\documentclass[%
 reprint,
nofootinbib,
 amsmath,amssymb,
 aps,
pre,
]{revtex4-2}

\usepackage{graphicx}
\usepackage{dcolumn}
\usepackage{bm}
\usepackage{hyperref}
\usepackage{subcaption}

\begin{document}

\preprint{APS/123-QED}

\title{Synchronization of Online Social Rhythms via Avatar Communications}
\thanks{Synchronization of Online Social Rhythms via Avatar Communications}%

\author{Masanori Takano}
\email{takano_masanori@cyberagent.co.jp}
 \affiliation{%
 Multi-disciplinary Information Science Center, CyberAgent, Inc.
}%

\date{\today}

\begin{abstract}
In this study, we consider users' online communication rhythms (online social rhythms) as coupled oscillators in a complex social network.
Users' rhythms may be entrained onto those of their friends, and macro-scale pattern of such rhythms can emerge.
We investigated the entrainment in online social rhythms and long-range correlations of the rhythms using an avatar communication dataset.
We indicated entrainment in online social rhythms to emerge if the strength of a new connection reaches a threshold.
This entrainment spread via densely-connected clusters.
Consequently, long-range correlations of online social rhythms extended to about 36\% of the network, although offline social life naturally restricts online social rhythms.
This research supports an understanding of human social dynamics in terms of systems of coupled oscillators.

\end{abstract}

\maketitle


\section{Introduction}

Many natural phenomena can be understood in terms of the synchronization of coupled oscillators in a network model~\cite{Arenas2008}, such as populations of bioluminescent fireflies~\cite{Buck1968}, rhythmic applause~\cite{Neda2000}, and crowd synchrony on bridges~\cite{Strogatz2005}.
Synchronization of coupled oscillators occurs when the connection between two oscillators exceeds a threshold~\cite{Kuramoto1984,Strogatz2000,Acebron2005}.
Global patterns of coupled oscillators, i.e., long-range correlations among systems of oscillators, emerge depending on the characteristics of network topologies, such as degree distributions, network radius, and clustering coefficients~\cite{Arenas2008,Rakshit2020}.

Human circadian rhythms can also be modeled as an oscillator.
These rhythms are not only coordinated by white light stimuli but also affected by social cues~\cite{Mistlberger2004}; for example, people coordinate their waking hours to attend work and school and to meet with friends.
We can regard humans' circadian rhythms as coupled oscillators in a complex social network~\cite{Aledavood2015,Astaburuaga2019}.

This should also be valid for online social rhythms, that is, the rhythms of online social activities, which also tend to provide information on peoples' physical rhythms in their everyday lives~\cite{Yokotani2021,Aledavood2022}.
People's communication activities are often driven by receiving online messages~\cite{Kaltenbrunner2008,Alakorkko2020}.
Users in an online communication application may coordinate their online social rhythms to communicate with their online friends.
That is, a user's social rhythm can be attracted to those of their friends.
Hence, online communication data enables us to observe global patterns of online social rhythms in online social networks.

Several studies have been conducted on the macroscopic behavior of circadian rhythms due to social interactions (social synchronization) in nonhuman organisms, e.g., ants~\cite{Fujioka2021}, fruit flies~\cite{Lone2011}, honey bees~\cite{Siehler2021}, and shorebirds~\cite{Bulla2016}.
These macroscopic behaviors describe population strategies such as role allotment and social coordination in the population.

Human communication dynamics are typically modeled as temporal networks, e.g., ~\cite{Kovanen2011,Aledavood2015,Yi-QingZhang2015,Aledavood2015a,Paranjape2017,Alakorkko2020,TakanoJPC2021}.
These dynamics include circadian and social rhythms~\cite{Aledavood2015,Aledavood2015a}.
However, relatively few works in the relevant literature have considered the relationship between humans' social synchronization of such rhythms and social networks~\cite{Astaburuaga2019}.
In this study, we investigate this relationship to understand human communication dynamics.
Additionally, the present work contributes to research on clinical psychology because disturbance of humans' circadian rhythms~\cite{Margraf2016,Astaburuaga2019,Crowe2020} and online social rhythms\cite{Lemola2011,Smarr2018,Yokotani2021,Yokotani2022cp} relate to exacerbations of mental health issues.

In this study, we first show that users' social rhythms are attracted to those of their friends, and the conditions of their relations.
Next, we demonstrate how similarities in social rhythms are associated with network structures.
Finally, we show the long-range correlations of similarities in social rhythms.

\section{Materials and Methods}
\subsection{Dataset}

\begin{figure}[t]
\centering
  \includegraphics[width=0.7\columnwidth]{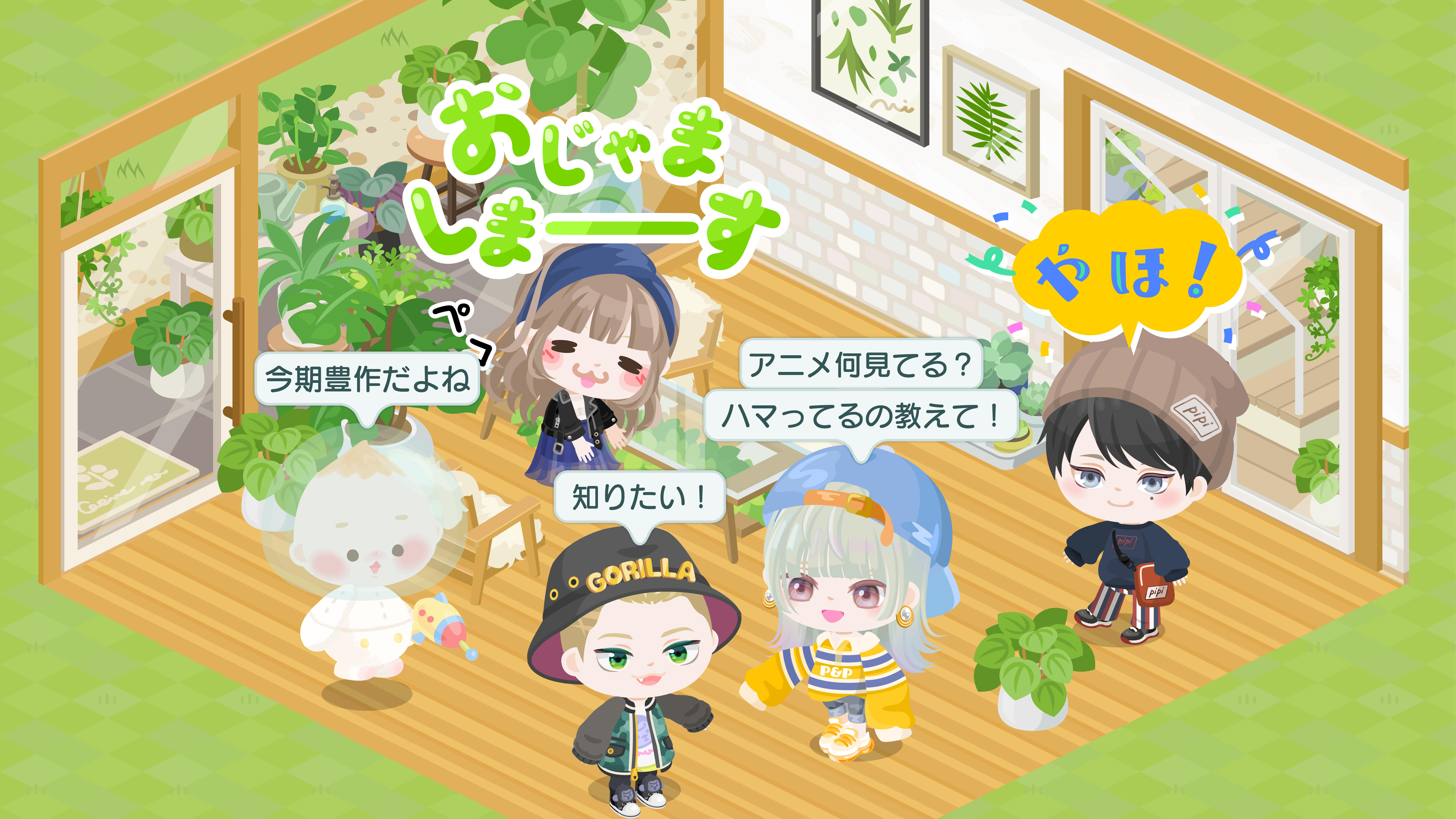}
  \caption{
  Players chat with each other via their avatars in Pigg Party.
  }~\label{fig_piggparty}
\end{figure}

We used a dataset of an avatar communication application called Pigg Party\footnote{\url{https://lp.pigg-party.com/}}, which was collected from March 1 to September 7, 2020, i.e., over the course of 27 weeks.
Pigg Party is a popular Japanese avatar communication application\footnote{A previous study reported that there were at least 550,000 active players over a six-month period~\cite{Yokotani2021_chb2}.
The authors excluded players who chatted with a specific user more than 100 times a day from their analysis.}.
In Pigg Party, players communicate using personalized avatars in virtual spaces (Fig.~\ref{fig_piggparty}).
Pigg Party players are typically women and young people.
In this application, 61\% of players were female, and 65\% were teenagers~\cite{MasanoriTakano2019}.

Players can synchronously communicate with each other in the game through their avatars in virtual spaces.
In addition to sending text messages, players can respond with dozens of avatar animations known as avatar actions.

Pigg Party offers private rooms as communication spaces for each player.
Players can enter a private room by
{\bf 1)} clicking the {\it enter} button shown in the profile window of the room's owner\footnote{Another player's profile window can be viewed if the player is an {\it acquaintance} or if they are in the same room.}, or
{\bf 2)} random entry can be initiated by a random entry mode.
In private rooms, players prefer talking with relatively few friends~\citep{MasanoriTakano2019}.

For each week, we sampled participants included in online social networks (discussed below in the section titled ``Online Social Networks'').
The mean number of participants each week was 65,229 (standard deviation: 5,087).

\subsection{Ethics Statement}
The Pigg Party application provider collects log data based on their terms of service\footnote{\url{https://lp.pigg-party.com/terms}} and privacy policy\footnote{\url{https://www.cyberagent.co.jp/way/security/privacy/}}.
All Pigg Party users accepted the terms of service and privacy policy, which allowed analysis of their behavioral data for service improvements and academic studies.
The data were pseudonymized, and identifying information was removed.

Quantitative data outputs are presented at an aggregate level, meaning no identifying information was included.

\subsection{Online Social Rhythms}

We constructed vectors of online social rhythms over a 24-hour cycle.
Each dimension of the vector expresses the activity of a given hour-long period (e.g., 1 o'clock, $\cdots$, 24 o'clock), i.e., 24 dimensions.

People's daily lives typically include various fluctuations that do not involve the 24-hour cycle~\cite{Aledavood2015}, such as single-day events, weekly events, or random events.
Therefore, we constructed the vector expressing a 24-hour cycle using usage time series data for each week with the fluctuations removed.

Previous studies~\cite{Yokotani2021,Yokotani2022cp} characterized online social rhythms using a 24-hour cycle factor based on discrete Fourier transforms.
This approach excludes detailed cycles.
For example, a 12-hour cycle factor of a user who communicates with online friends in the morning and evening is more significant than their 24-hour cycle factor.

We extended this approach to construct detailed online social rhythms.
We used eleven cycle factors, which were calculated by 24 hours divided by integers $i$ ($i=1, 2, \cdots, 11$), i.e., 24, 12, 8, 6, 4.80, 4, 3.43, 3, 2.67, 2.40, and 2.18 hours.
That is, online social rhythm vectors were constructed based on eleven bandpass-filtered frequencies.

We describe the procedure used to construct online social rhythms as follows.

First, we constructed time series from users' usage minutes for each hour in a week (Fig.~\ref{fig_si_week_usage}).
Second, we conducted discrete Fourier transforms of the users' usage time series (their power spectrums are shown in Fig.~\ref{fig_si_spector}).
Third, we extracted the eleven frequency elements.
Finally, we calculated the inverse Fourier transforms of the filtered elements.
We then normalized the results to obtain an online social rhythm vector for each user for each week (Fig.~\ref{fig_si_rhythm}).

A larger product of the rhythm vectors for each hour indicates that two users are more likely to use the system simultaneously.
Therefore, we used the inner product of rhythm vectors as social rhythm similarity.
This value is equivalent to the cosine similarity owing to the normalization of the rhythm vectors.

\subsection{Online Social Networks}

We constructed online social networks to analyze the relationships between social rhythm similarity and social connections.
We used the logs of users' visits to others' private rooms.
We assumed an edge between a visitor and the owner of a room, as in previous work~\cite{Yokotani2021,takano_icwsm2022,yokotani_tele2022}, because visitors tend to be the owner's friends~\cite{MasanoriTakano2019}.
The dwelling time (sec) at over a week were used to represent the weights of the edges ($w$).
We aggregated the dwelling time on a weekly basis.
The mean power law coefficients of degree distributions of these networks  (mean: $4.80$ and standard deviation: $0.30$) were larger than those of general social networks~\cite{Barabasi2016}.
That is, the features of these networks were more similar to random networks~\cite{Barabasi2016}, e.g., with relatively few hub nodes.

Although this network does not directly represent the communication network, the model was positively associated with the actual communication network.
We used this network as an approximation because data on the actual communication paths between people are strictly restricted based on constitutionally guaranteed secrecy of communication in Japan.

We used a null network to evaluate the effects of social rhythm similarity.
This refers to randomized networks in which all edges were randomly reattached via ``random combination.''

\section{Results}


\begin{figure}[t]
\centering
  \includegraphics[width=0.7\columnwidth]{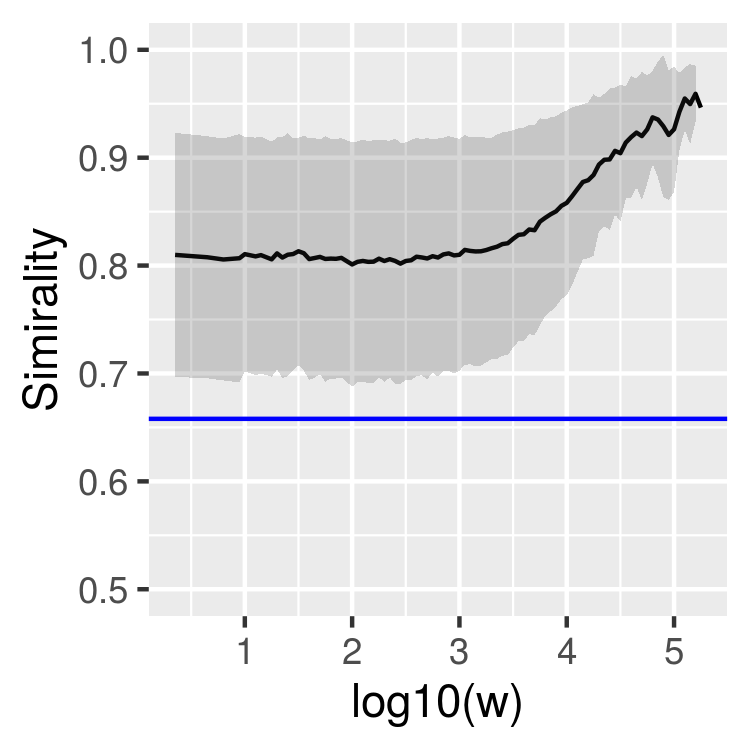}
  \caption{
  Nonlinear relationships between social rhythm similarities and edge weights.
  The semi-transparent ribbon represents the standard deviation in the following figures.
  The blue line shows the mean similarity of randomly reattached users.
  }~\label{fig_sim_staytime}
\end{figure}

\begin{figure}[t]
\centering
  \includegraphics[width=1\columnwidth]{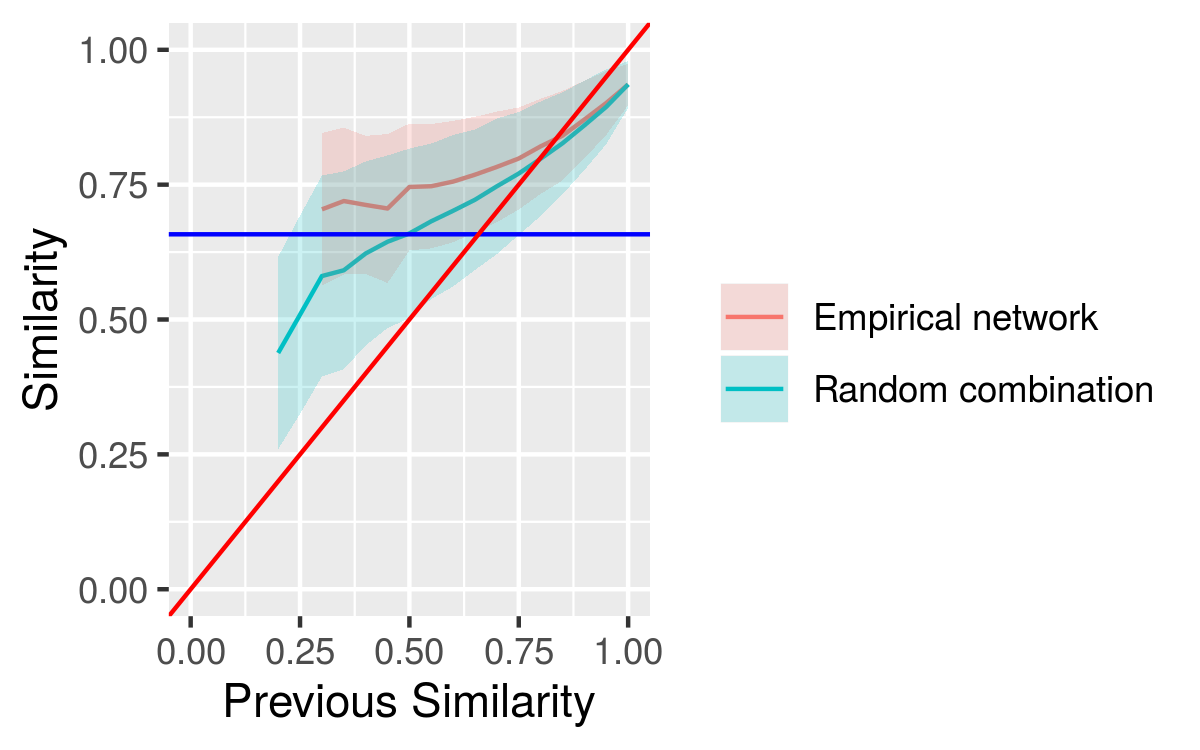}
  \caption{
  Changes of similarities from before connecting users to afterwards in terms of empirical networks and random combinations.
  On the red line, the similarities are unchanged.
  }~\label{fig_change_sim}
\end{figure}
\begin{figure}[t]
\centering
  \includegraphics[width=1\columnwidth]{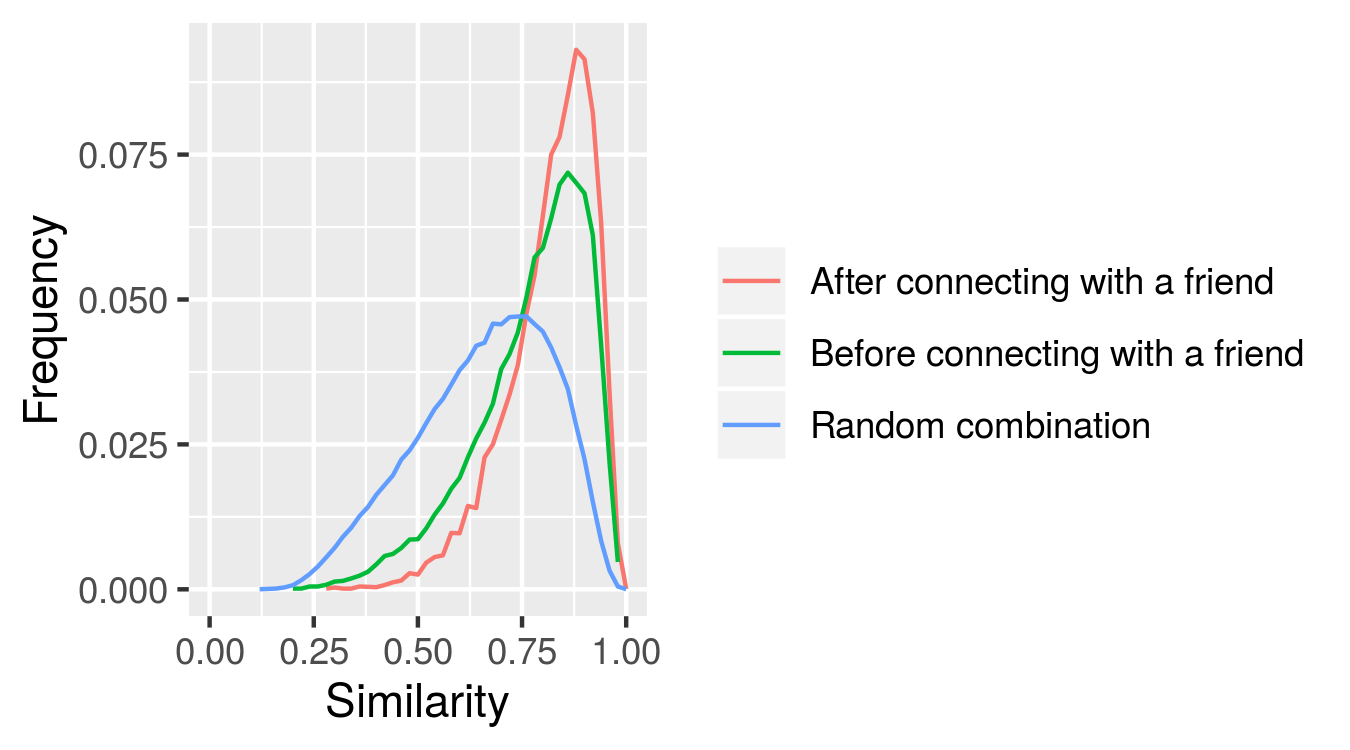}
  \caption{
  Distributions of similarities of users before and after connecting and random combinations
  }~\label{fig_sim_dist}
\end{figure}

\begin{figure}[t]
\centering
  \includegraphics[width=1\columnwidth]{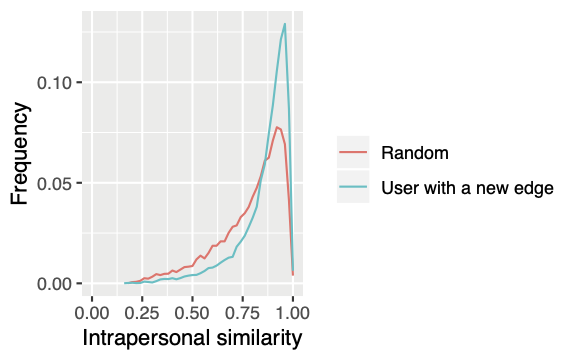}
  \caption{
  Intrapersonal similarities between a user's rhythms before and after connecting with friends
  }~\label{fig_rhythm_stability}
\end{figure}

\begin{figure}[t]
\centering
  \includegraphics[width=0.5\columnwidth]{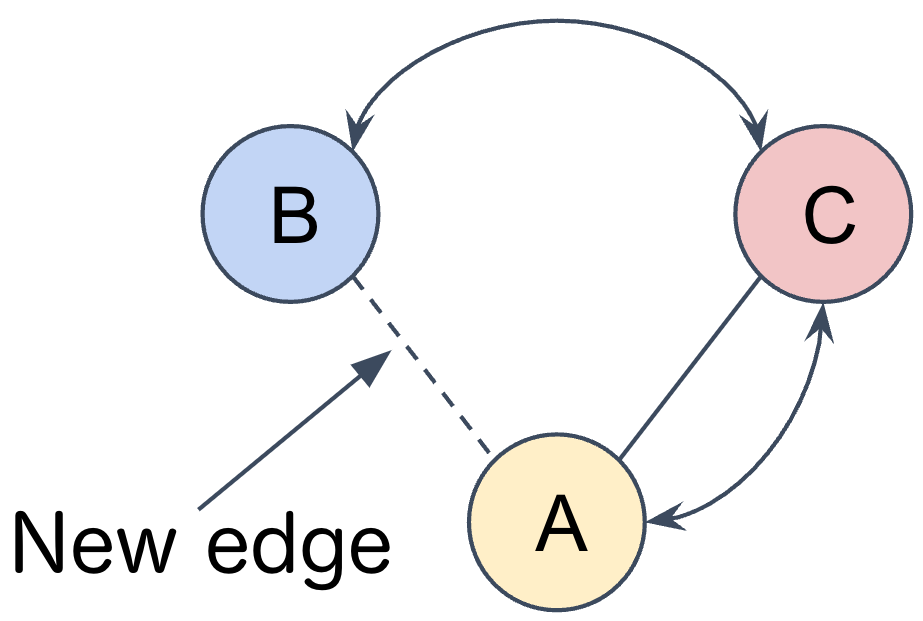}
  \caption{
  Overview of analysis of social rhythm changes in triadic relations
  }~\label{fig_new_edge}
\end{figure}

\begin{figure}[t]
\centering

  \begin{minipage}[b]{1\linewidth}
    \centering
\includegraphics[width=1\columnwidth]{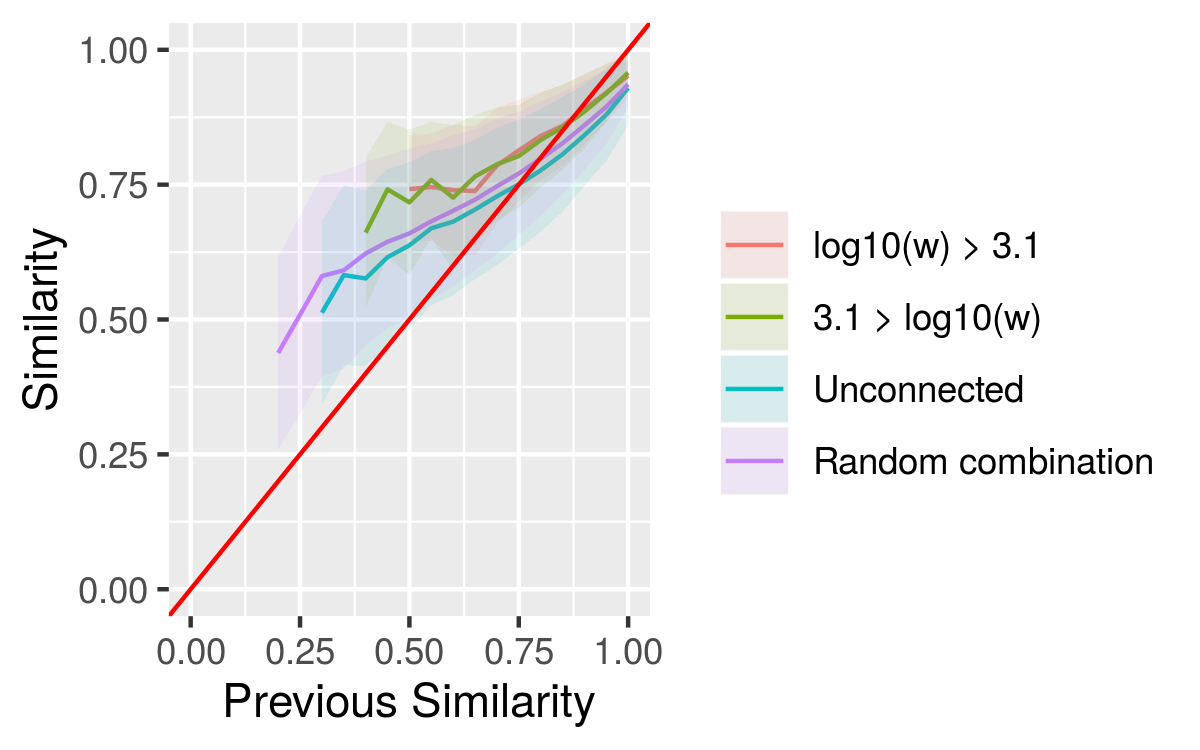}
      \subcaption{A-C}
  \end{minipage}
  \hspace{10mm}
  \begin{minipage}[b]{1\linewidth}
    \centering
  \includegraphics[width=1\columnwidth]{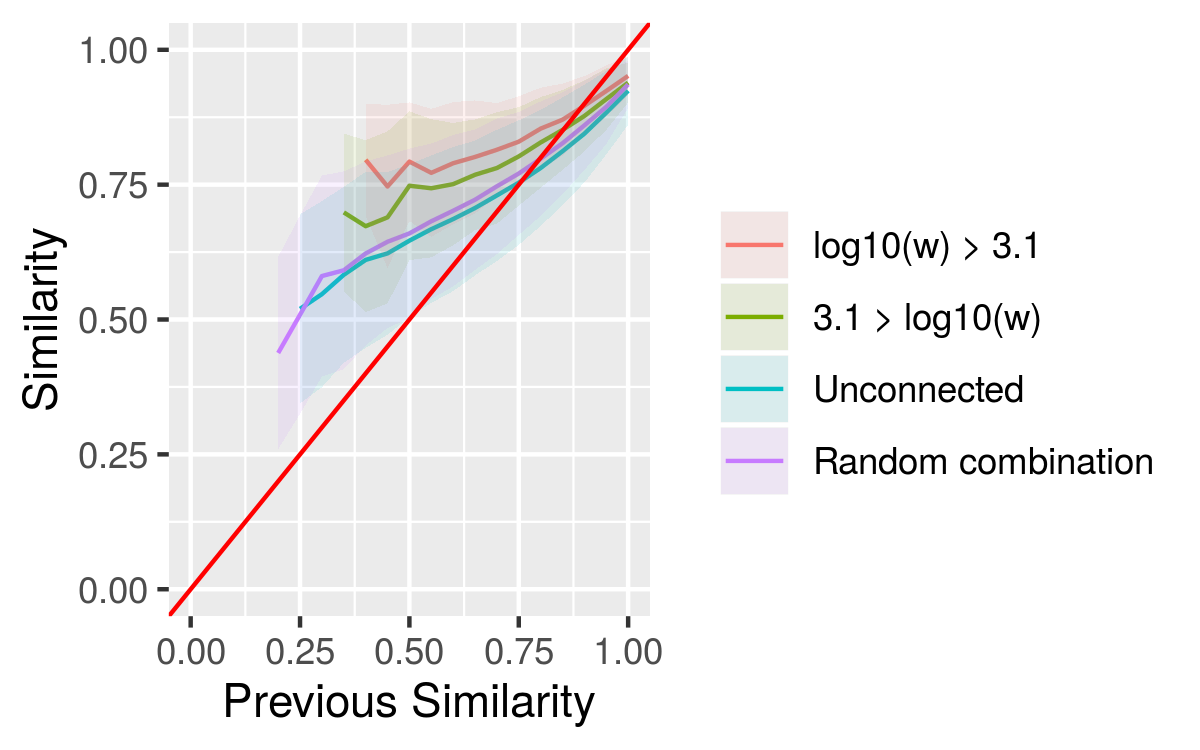}
    \subcaption{B-C}
  \end{minipage}
  \caption{
  Similarity changes of connection A--C and B--C when connections A--B arose (Fig.~\ref{fig_new_edge}).
  }~\label{fig_change_sim_3rd}
\end{figure}

\begin{figure}[t]
\centering
  \includegraphics[width=0.7\columnwidth]{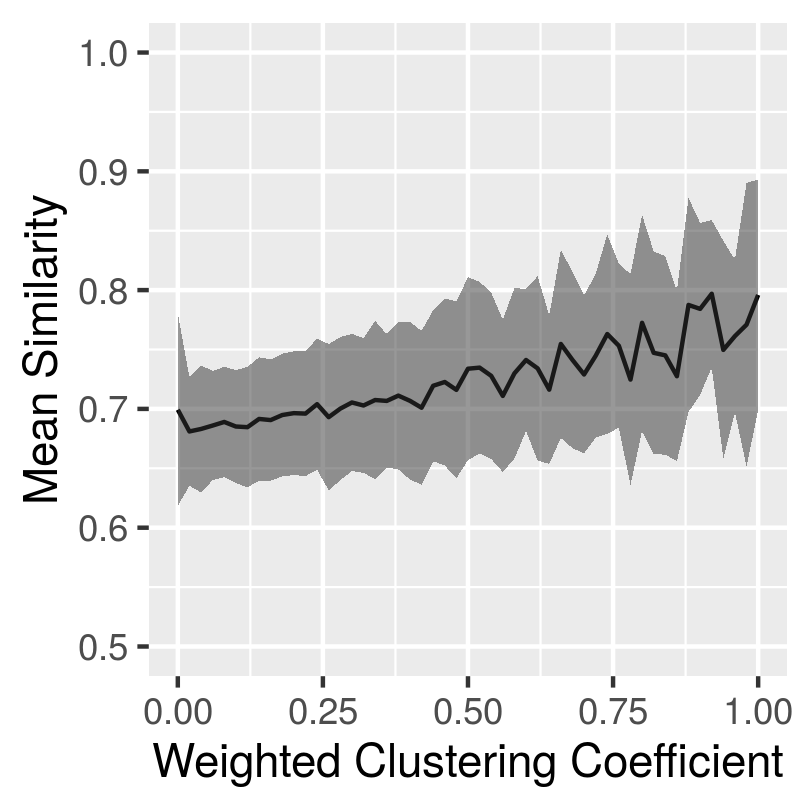}
  \caption{
  Relationships between mean similarities of community members and weighted clustering coefficients~\cite{Clemente2018} of the communities.
  Communities were detected using the Infomap algorithm~\cite{Rosvall2008} on each week's social network.
  }~\label{fig_cc}
\end{figure}

\begin{figure}[t]
\centering
    \centering
    \includegraphics[width=1\linewidth]{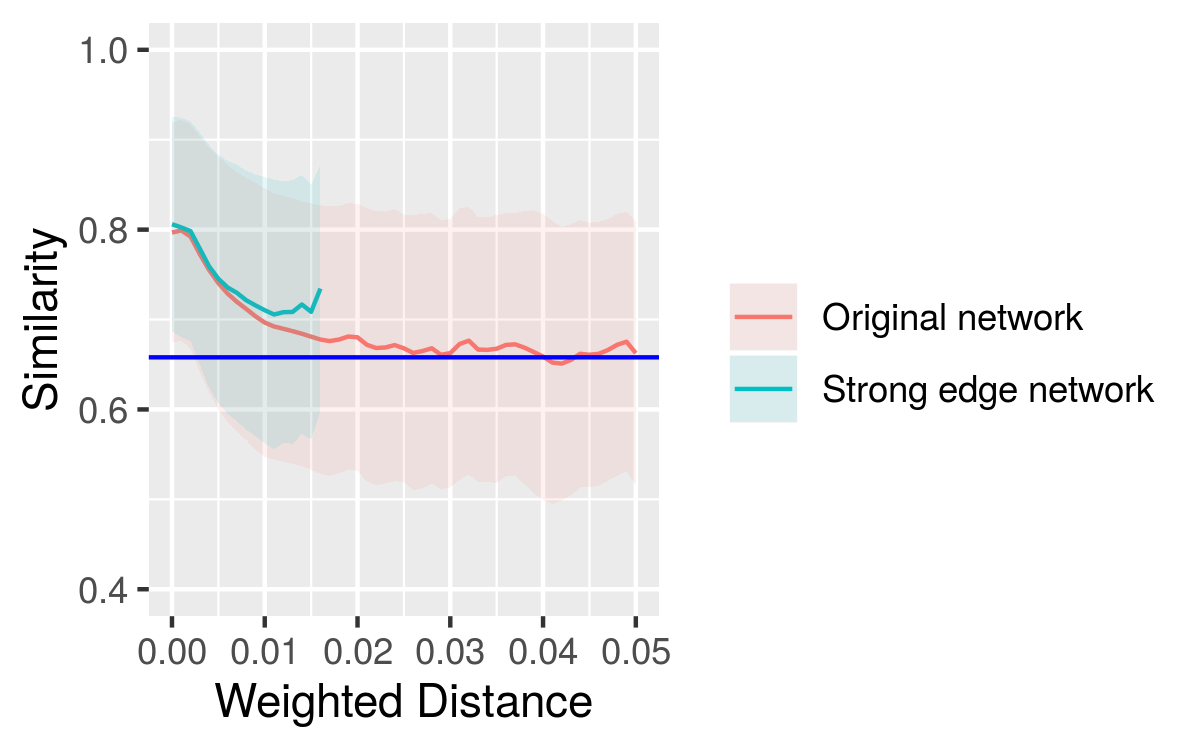}
  \caption{
  Long-range correlations of social rhythms in weighted networks.
  We used $1/w$ as distances between nodes.
  The original networks include all edges.
  The strong edge networks include only strong edges ($\log_{10} w \geq 3.1$).
  The maximum distance of the strong edge networks was shorter than that of the original networks because strong edge networks were separated by excluding weak edges.
  }
  \label{fig_distance}
\end{figure}

\subsection{Edge Weights and Social Rhythm Similarity}
We found the nonlinear relationship between social rhythm similarities and edge weights (Fig.~\ref{fig_sim_staytime}).
We used the similarities between randomly reattached users as baseline (mean: $0.658$ and standard deviation: $0.161$).
The similarities increased with $w$ as a threshold $\log_{10} w > 3.1$ (about 21 minutes), i.e., social rhythm similarities were proportional to $\log_{10} w$.
Even connections of $\log_{10} w \leq 3.1$ showed more similar rhythms than the baseline.
The difference between them was $0.149^{***}$ (Welch's t-test).
This seemed to have occurred simply because users with similar rhythms tended to meet incidentally.

\subsection{Entrainment of Social Rhythms}
Fig.~\ref{fig_change_sim} shows the changes of similarities from before connecting users to afterwards in terms of empirical networks and random combinations.
The values of random combinations were calculated by combining randomized networks of a given week and the previous week.
Users' social rhythms became similar when they connected, i.e., their rhythms were entrained, compared with randomly reattached users (green line) and the baseline (blue line).
The difference between the empirical network and the baseline was $0.052^{**}$ (Welch's t-test) when the previous similarity was low (i.e., it was in the range $[0.30, 0.35]$).
In contrast, similarity varied less between randomly reattached users than the baseline when the previous similarity was low.
Thus, the similarity between unconnected users increased over that of random combinations and the baseline even if their previous similarity was low.

The rhythms of users who connected in the next week, were similar before connecting, compared with random combinations (Fig.~\ref{fig_sim_dist}).
Note that users who had different rhythms sometimes connected.
Their rhythms were more similar after connecting than that before doing so.
The difference was $0.046^{***}$ (Welch's t-test).

The entrainment of social rhythm due to a new edge stabilized their rhythms.
Fig.~\ref{fig_rhythm_stability} shows the distributions of intrapersonal similarities between a user's rhythms before and after connecting with friends in the game.
The intrapersonal similarities of users with a new edge (mean: $0.859$ and standard deviation: $0.125$) were higher than those of randomly sampled users (mean: $0.790$ and standard deviation: $0.160$).
The mean difference was $0.0693^{***}$ (Welch's t-test).

\subsection{Entrainment of Social Rhythms in Triadic Relations}

We analyzed changes in social rhythm similarity in triadic relations in which a new connection emerged (Fig.~\ref{fig_new_edge}).
Fig.~\ref{fig_change_sim_3rd} shows the changes in the similarity of Connections A--C and B--C when connection A--B arose.
We considered three conditions to model the relation between B and C, including a strong edge ($\log_{10}w>3.1$), a weak edge ($3.1\leq \log_{10}w$), and unconnected.

As a result, the similarities of A--C and B--C increased when A--B connections arose if B--C had connected.
That is, a social rhythm change arising from a new connection did not spread through existing edges when their topology was linear (the unconnected condition).
However, if their topology was clustered, their rhythm similarities increased.
Consequently, the members of dense communities with high clustering coefficients showed high similarities (Fig.~\ref{fig_cc}).
\subsection{Long-range Correlations}

The relationships between rhythm similarities and weighted distances used to analyze the long-distance correlations are shown in Fig.~\ref{fig_distance}.
Rhythm similarities exponentially decreased with distances.
The distance of the similarity ($0.720$), which was 10\% larger than the baseline, was $0.007$.
The distance $0.007$ reached 35.8\% of networks.
The weighted paths with the distance $0.007$ required about 4 or 5 hops (Fig.~\ref{fig_si_hop_dist}).
This tendency was the same for unweighted networks, i.e., those in which all edge weights were $1$ (Fig.~\ref{fig_si_uw_distance}).

Networks excluding weak edges (strong edge networks) exhibited similar results to original networks, although, in general, excluding edges stretched distances between nodes.
This means that strong edges were dominant in synchronizations of social rhythm.

\section{Discussion}

We showed that users' online social rhythms entrain to those of their friends, mutually, using a dataset from an avatar communication application.
That is, as social cues coordinate circadian rhythms~\cite{Mistlberger2004}, online social cues coordinate online social rhythms.
People's social rhythms naturally fluctuate to some extent~\cite{Aledavood2015}.
Although online social rhythms in our dataset also exhibited such fluctuation, they tended to stabilize with the entrainment of social rhythms.
The entrainment emerged when connections between users reached a certain threshold.
This relationship between two oscillators is consistent with theoretical studies~\cite{Kuramoto1984,Strogatz2000,Acebron2005} in which synchronization of coupled oscillators occurred when their connection exceeded a threshold.
At the threshold and above, the rhythm similarities of two nodes proportionally increase with $\log_{10} w$.
Consequently, the long-range correlation was dominated by connections with more substantial edges than the threshold.

When two people who had shared a friend connected, the rhythms of the three also synchronized.
In contrast, if two people had not shared a friend, a new connection between them did not affect a third person who was a friend of either of them.
In other words, the social contagion of online social rhythms requires direct connections.
As a result, communities with high clustering coefficients exhibited high similarity.
This suggests that online social rhythms spread via densely-connected clusters.
This phenomenon is consistent with theoretical results in which high clustering coefficients facilitate synchronization of oscillators~\cite{Arenas2008}.

Theoretical models show long-range correlations over the entire extent of a network~\cite{Ichinomiya2004,Arenas2008,Rakshit2020}.
This is because a robust synchronization emerged on a densely-connected cluster (core), and the synchronization entrained neighbor nodes~\cite{Ichinomiya2004}.
As a result, all oscillators synchronized.
Similarly, online social rhythm also showed a synchronization across density clusters.
Consequently, long-range correlations of online social rhythm reached about 36\% of a social network, although offline social life naturally restricts online social rhythms.

These findings highlight the importance of social rhythms on human communication dynamics.
These dynamics have typically been studied as temporal communication networks based on detailed communication events, e.g., sending messages~\cite{Kovanen2011,Aledavood2015,Yi-QingZhang2015,Aledavood2015a,Paranjape2017,TakanoJPC2021}.
In contrast, our approach of modeling and analyzing communicating users as coupled oscillators in a social network provided some novel insight from a different perspective.

Studies have shown that disturbance of circadian rhythms relates to exacerbation of mental health issues~\cite{Margraf2016,Crowe2020}.
Some studies reported that disturbances in online social rhythms were also associated with exacerbation of mental health issues~\cite{Lemola2011,Smarr2018,Yokotani2021,Yokotani2022cp}.
Interpersonal and social rhythm therapy (IPSRT)~\cite{Frank2005} improves mental health by adjusting circadian rhythm.
Our findings suggest that adjusting not only a patient's social rhythm but also that of their friends may be effective, because densely-connected clusters' social rhythms seem to be robust.
We also found that communicating with a new friend can stabilize social rhythms due to an entrainment with the friend's social rhythm.

\section{Summary}

Using an avatar communication dataset, we have modeled the social synchronization of online social rhythms and their long-range correlations.
Such spreading synchronization, i.e., emergent long-range correlations, seems to require close and densely-connected communities.
The results of this study provide new evidence of online social synchronization, an essential feature of temporal social networks~\cite{Aledavood2015,Aledavood2015a,Alakorkko2020}, and contribute to the mathematical modeling of temporal social networks.
Our findings can also contribute to the literature on clinical psychology focusing on social rhythms, such as IPSRT.

\bibliographystyle{apsrev}

\section*{Acknowledgements}
We are grateful to Dr. Soichiro Morishita and Mr. Makoto Takeuchi at CyberAgent, Inc., whose comments and suggestions were very valuable throughout this study.

\newpage

\appendix
\setcounter{equation}{0}
\setcounter{figure}{0}
\setcounter{table}{0}

\renewcommand{\figurename}{Fig.}
\renewcommand{\thefigure}{A\arabic{figure}}
\renewcommand{\tablename}{Table.}
\renewcommand{\thetable}{A\arabic{table}}

\renewcommand{\theequation}{A\arabic{equation}}


\begin{figure*}[!h]
\LARGE{Appendix}
\hspace{10mm}

  \begin{minipage}[b]{0.49\linewidth}
\centering
  \includegraphics[width=1\columnwidth]{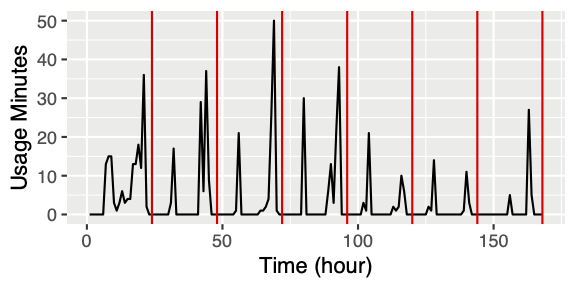}
  \caption{A sample users' time series of usage minutes of each hour in a week.
  The red line shows 24 o'clock of each day.
  }~\label{fig_si_week_usage}
  \end{minipage}

  \begin{minipage}[b]{1\linewidth}
    \begin{minipage}[b]{0.49\linewidth}
    \centering
    \includegraphics[width=0.7\linewidth]{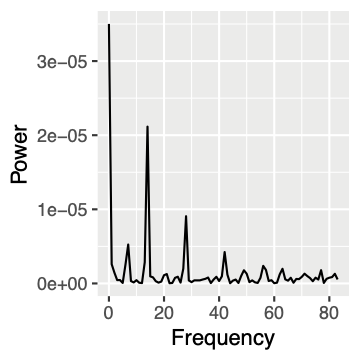}
    \subcaption{Sample user}
  \end{minipage}
  \begin{minipage}[b]{0.49\linewidth}
    \centering
    \includegraphics[width=0.7\linewidth]{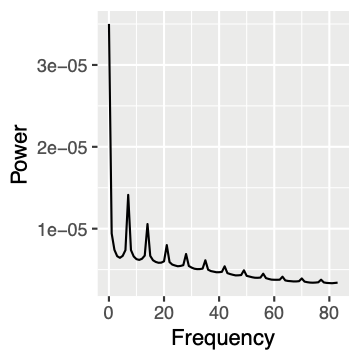}
    \subcaption{Mean of all users}
  \end{minipage}
  \caption{Power spectrums of the sample user in Fig.~\ref{fig_si_week_usage} and the mean of all users in a week (July 26, 2020, to August 1, 2020).
  The peaks in these figures are frequency elements by which 24 hours were exactly divisible, i.e., these periods were 24, 12, 8, 6, 4.80, 4, 3.43, 3, 2.67, 2.40, and 2.18 hours long.}
  \label{fig_si_spector}
  \end{minipage}

  \begin{minipage}[b]{0.49\linewidth}
  \centering
  \includegraphics[width=0.5\columnwidth]{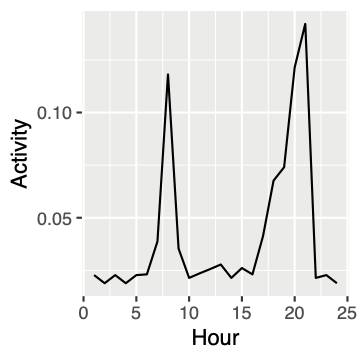}
  \caption{Social rhythm of the sample user in Fig.~\ref{fig_si_week_usage}
  }~\label{fig_si_rhythm}
  \end{minipage}

\end{figure*}

\begin{figure*}[!h]
  \begin{minipage}[b]{1\linewidth}
    \begin{minipage}[b]{0.48\linewidth}
    \centering
\includegraphics[width=1\columnwidth]{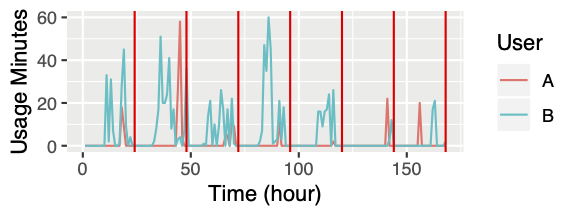}
      \subcaption{Before connecting with a friend}
  \end{minipage}
  \begin{minipage}[b]{0.48\linewidth}
    \centering
  \includegraphics[width=1\columnwidth]{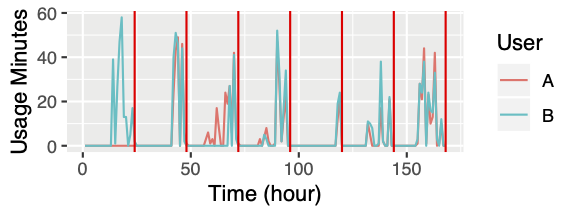}
    \subcaption{After connecting with a friend}
  \end{minipage}
  \caption{Two users' time series of usage minutes for each hour in a week
  }~\label{fig_si_u12_ts}
  \end{minipage}

  \begin{minipage}[b]{1\linewidth}
    \begin{minipage}[b]{0.48\linewidth}
    \centering
\includegraphics[width=0.8\columnwidth]{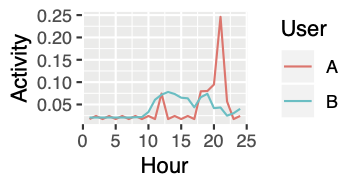}
      \subcaption{Before connecting with a friend}
  \end{minipage}
  \begin{minipage}[b]{0.48\linewidth}
    \centering
  \includegraphics[width=0.8\columnwidth]{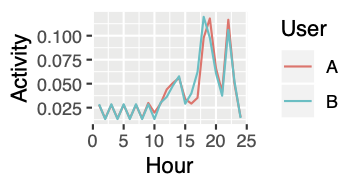}
    \subcaption{After connecting with a friend}
  \end{minipage}
  \caption{Two users' social rhythms
  }~\label{fig_si_u12_rhythm}
    \end{minipage}

  \begin{minipage}[b]{0.45\linewidth}
    \centering
    \includegraphics[width=0.7\linewidth]{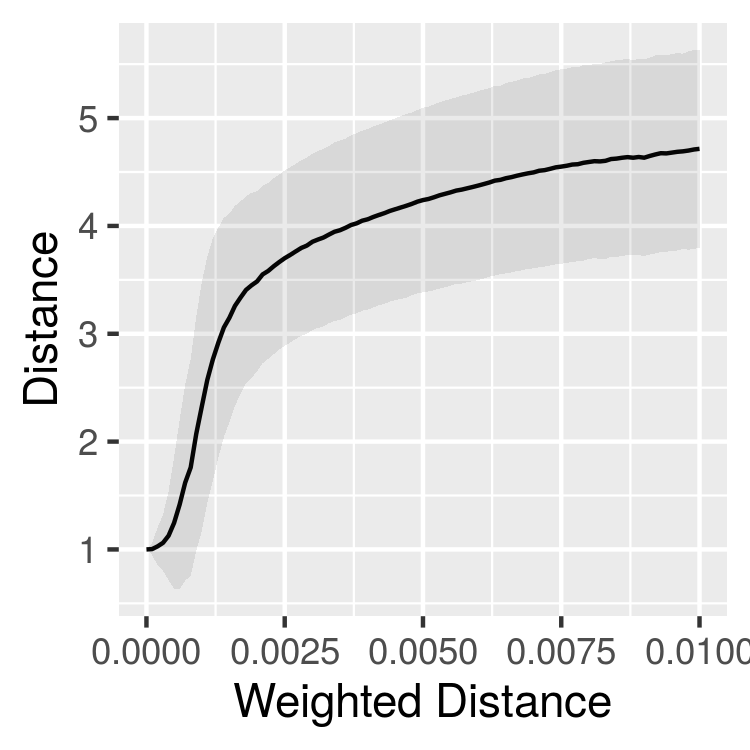}
  \caption{Relationships between weighted and unweighted distances
  }~\label{fig_si_hop_dist}
\end{minipage}
  \begin{minipage}[b]{0.45\linewidth}
    \centering
    \includegraphics[width=0.7\linewidth]{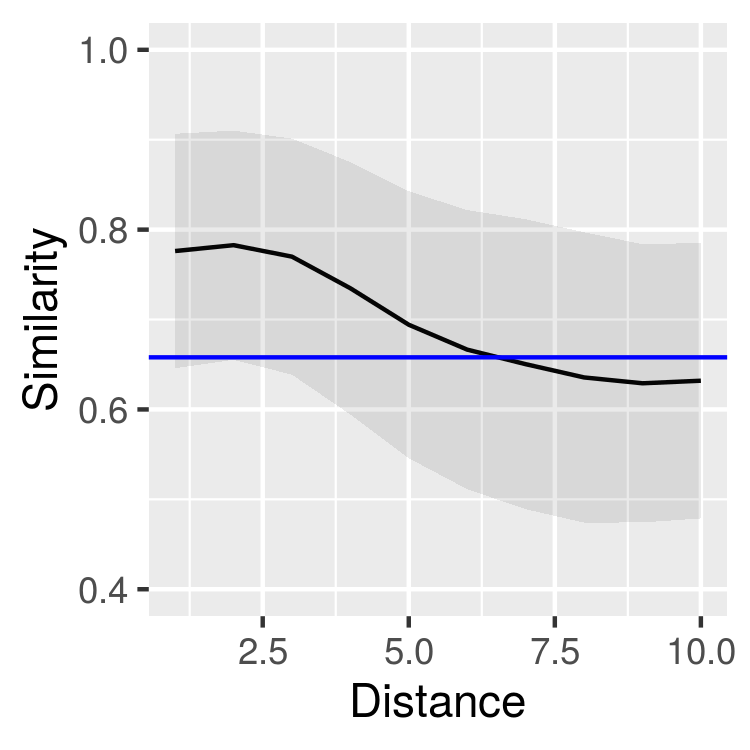}
  \caption{
  Long-range correlations of social rhythm in unweighted networks.
  We regarded all distances between nodes as 1.
  The distances of the similarity ($0.723$), which was 10\% larger than the base line, was $4$.
    The distance $4$ reached up to 33.1\% of the network.
  }~\label{fig_si_uw_distance}
    \end{minipage}
\end{figure*}

\end{document}